\documentclass[11pt,a4paper]{article}
\usepackage{geometry}
\usepackage{graphics}
\usepackage{setspace}
\usepackage[scaled=0.9]{helvet}
\usepackage{graphicx}
\usepackage{subfig}
\usepackage{amssymb}
\usepackage{rotating}
\usepackage{arydshln}
\usepackage{multicol}
\usepackage{abstract}
\usepackage{ucs}
\usepackage[utf8x,utf8]{inputenc}
\usepackage[T1]{fontenc}
\usepackage[swedish, english]{babel}
\usepackage{fancyhdr}

\usepackage{slashed}

\begin{document}
\bibliographystyle{acm}
\pagestyle{fancy}
\cfoot{\thepage}
\renewcommand{\abstractname}{}

\title{\fontfamily{phv}\selectfont{\huge{\bfseries{Toward particle-level filtering of individual collision events at the Large Hadron Collider and beyond}}}}
\author{
{\fontfamily{ptm}\selectfont{\large{Federico Colecchia}}}\thanks{Email: federico.colecchia@brunel.ac.uk}\\
{\fontfamily{ptm}\selectfont{\large{{\it Brunel University, London, Uxbridge, UB8 3PH, United Kingdom}}}}
}
\date{}
\maketitle
\begin{onecolabstract}
Low-energy strong interactions are a major source of background at hadron colliders, and methods of subtracting the associated energy flow are well established in the field. Traditional approaches treat the contamination as diffuse, and estimate background energy levels either by averaging over large data sets or by restricting to given kinematic regions inside individual collision events. On the other hand, more recent techniques take into account the discrete nature of background, most notably by exploiting the presence of substructure inside hard jets, i.e. inside collections of particles originating from scattered hard quarks and gluons. However, none of the existing methods subtract background at the level of individual particles inside events. We illustrate the use of an algorithm that can enable particle-by-particle background discrimination at the Large Hadron Collider, and we envisage this as the basis for a novel event filtering procedure upstream of the official jet reconstruction pipelines. Our hope is that this new technique will improve physics analysis when used in combination with state-of-the-art algorithms in high-luminosity hadron collider environments.
\end{onecolabstract}

\begin{multicols}{2}
{\bf Keywords:}
29.85.Fj; High Energy Physics; Particle Physics; Large Hadron Collider; LHC; background discrimination; mixture models; latent variable models; sampling; Gibbs sampler; Markov Chain Monte Carlo; Expectation Maximisation; Multiple Imputation; Data Augmentation.

\section{Introduction}
\label{intro}

Strong interactions described by Quantum Chromodynamics (QCD) play a major role at hadron collider experiments such as those at the Large Hadron Collider (LHC) at CERN, where the highest-energy proton beams available worldwide are collided. The higher event multiplicities and background rates as compared to previous experiments have an impact on physics analysis, and place even stronger requirements on background subtraction than they did in the past.

In particular, the energy flow associated with soft, i.e. low-energy, QCD interactions is an important background at the LHC. Pileup, i.e. particles originating from proton-proton collisions that are not the one of interest but that nonetheless contribute to the same triggered event, is an issue to a number of LHC analyses, and its impact is going to become more and more relevant as the instantaneous luminosity of the accelerator is increased. 

The high pileup rates that are foreseen at upgraded LHC scenarios can significantly affect searches for new heavy particles in final states containing missing transverse momentum, $\slashed{p}_T$ \footnote{
Missing transverse momentum, $\slashed{p}_T$, is the observed momentum imbalance inside an event measured on a plane perpendicular to the beam direction.}, 
as well as the analysis of channels containing 
hard jets, i.e. collections of particles originating from scattered hard quarks and gluons. 
In fact, jet energy correction 
has a direct impact on the quality of the reconstructed jet objects that are ultimately used for analysis (see e.g. \cite{CMS_calib, CMS_calib_2}). 

In addition to pileup, when a hard parton scattering from a proton-proton collision takes place, additional particles are also produced by the Underlying Event, i.e. by 
interactions between the proton beam remnants and by 
multiple parton interactions. Moreover, particles can generate additional energy flow in the form of initial-state radiation prior to the hard scattering. All of these effects can have a notable impact on physics analysis and are carefully taken into account at the experiments during reconstruction and calibration. 

\section{The state of the art}
\label{review}

Methods of subtracting soft QCD background associated with pileup and Underlying Event at hadron colliders are well established. 
Traditional techniques treat the contamination as diffuse, and estimate a background momentum contribution that is then subtracted from the total momentum of the hard jets of interest. 
Some of these methods rely on high-statistics samples, e.g. 
Minimum Bias, dijet, and Drell-Yan data 
(see e.g. \cite{CMS_calib_2}). 
Given the way these methods work, the estimated background energy contribution is typically averaged over many events, and event-to-event background fluctuations 
are therefore neglected.

Following the introduction of the notion of jet area \cite{jet-area}, which provides a measure of the susceptibility of reconstructed jets to soft QCD energy flow, the focus has shifted toward event-by-event estimation of the background momentum density. With jet area-based methods, the quantity that is subtracted from the total momentum of the hard jets is 
proportional to an event-level estimate of the background momentum density as well as to the area of the jet of interest. This takes into account possible event-to-event variations of the soft QCD energy flow. 

However, since the estimated momentum density is typically averaged over all pileup jets in the event, 
this approach still neglects background fluctuations inside individual events. 
%
%
Nonetheless, 
the amount of soft QCD contamination can be different in different jets due to the quantum nature of the underlying physics. 
Subtracting background in a kinematics-dependent way can partially address this issue, although jet area-based methods were ultimately not developed to describe such effects. 

A more recent approach 
exploits the presence of substructure inside jets. Jet grooming techniques 
are being used at the LHC 
to reject soft QCD contamination inside jets \cite{jet-grooming-01,jet-grooming-02,jet-grooming-03}. As opposed to treating the low-energy background as diffuse, these methods exploit the presence of substructure that is often associated with the hierarchical composition of the jets. 
Such methods have proven particularly useful, especially in combination with jet-vertex association techniques that map individual track jets to putative primary interaction vertices. 

\section{A candidate new approach}
\label{gibbshep}

Despite the wealth of techniques available and the effectiveness they have so far demonstrated, 
none of the existing methods use information at the finest-grained level, 
i.e. at the level of individual particles inside events. We elaborate on the possibility to use our algorithm \cite{gibbshep2, gibbshep} 
to implement  
a novel event filtering procedure to reject soft QCD background from individual LHC events particle by particle. We suggest that individual particles inside events can be mapped to a signal hard scattering as opposed to soft QCD background on a probabilistic basis, thereby taking into account the effect of 
fluctuations on the shapes of particle-level probability density functions (PDFs). 
This can be particularly useful with reference to neutral particles, which in general cannot be easily associated with the primary interaction vertex. 

%
%




\section{The algorithm}
\label{algo}


We recently presented a Markov Chain Monte Carlo algorithm that makes it possible to assign individual particles inside events a probability for them to originate from a hard parton scattering as opposed to soft QCD interactions. 
We showed results on Monte Carlo data sets comprising a total number of particles 
in the range $\sim1300\div 1600$, 
corresponding to $gg\rightarrow t\bar{t}$ at $\sqrt{s}=14~\mbox{TeV}$ superimposed with Minimum Bias. 
Our algorithm 
makes it possible to estimate the effect of fluctuations on the shapes of signal and background PDFs at the particle level. 
We are here 
discussing the possibility of 
using 
it 
to implement a particle-level filtering procedure for individual LHC events upstream of the official jet reconstruction pipelines. 

The algorithm processes a given collection of particles that is assumed to be a mixture comprising particles originating from a signal hard 
scattering as well as particles associated with soft QCD background. It samples iteratively from a Bayesian posterior PDF that encodes information as to which particles are more likely to originate from either process, discriminating based on  
signal and background particle-level kinematics. In particular, with reference to individual particle pseudorapidity\footnote{
Particle pseudorapidity, $\eta$, is a kinematic variable that is related to the particle polar angle, $\theta$, in the laboratory frame, and which is given by $\eta=-\log(\tan(\theta/2))$.
}, $\eta$, the distribution of particles originating from a hard quark or gluon scattering is typically more biased toward zero, i.e. the corresponding particles are more ``central'' in the detector as compared to particles associated with soft QCD interactions.


The statistical model is a convex combination of particle-level PDFs corresponding to the hard scattering and to soft QCD background: $\alpha_0 f_0(\eta, p_T) + \alpha_1 f_1(\eta, p_T)$. The quantity $\alpha_0$ ($\alpha_1$) is the fraction of background (signal) particles, and $f_0$ ($f_1$) is the background (signal) PDF. In practice, in the study described in \cite{gibbshep2}, most of the discrimination power comes from the $\eta$ distributions. The PDFs $f_j$ are estimated by regularising $\eta$ histograms based on spline interpolation of the bin contents. This provides the statistical model with the flexibility required to describe generic deviations of the PDF shapes from those of the corresponding control sample templates due to fluctuations. The symbol $\varphi_j$ will be used throughout to refer to such estimates of the PDFs $f_j$. The pseudocode is given below, $v^{(t)}$ referring to the value of variable $v$ at iteration $t$.

\begin{enumerate}
\item {\bf Initialization:} Set $\underline{\alpha}^{(0)}=\{\alpha_j^{(0)}\}_j$, $j=0,1$, and obtain estimates $\varphi_j^{(0)}$ of the subpopulation PDFs $f_j$ by regularising the corresponding 
distributions from a high-statistics control sample. 
\item {\bf Iteration $t$:}
\begin{enumerate}
\item Generate the ``allocation variables" $z_{ij}^{(t)}$, for all particles $i=1,...,N$, and $j=0,1$, based on the conditional probabilities $P(z_{ij}^{(t)}=1 | \alpha_j^{(t-1)}, \varphi_j^{(0)}, x_i)=\alpha_j^{(t-1)}\varphi_j^{(0)}(x_i)/(\alpha_0^{(t-1)}\varphi_0^{(0)}(x_i)+\alpha_1^{(t-1)}\varphi_1^{(0)}(x_i))$. The quantity $z_{ij}^{(t)}$ equals 1 when observation $i$ is mapped to distribution $j$ at iteration $t$, and 0 otherwise.
\item Map individual particles to signal or background based on $z_{ij}$, and set $\alpha_0^{(t)}$ ($\alpha_1^{(t)}$) to the fraction of particles mapped to background (signal) at iteration $t-1$.
\end{enumerate}
\end{enumerate}


As described in \cite{gibbshep2}, the algorithm was inspired by the Gibbs sampler \cite{geman}, and its development was influenced by a number of statistical techniques including Expectation Maximisation \cite{EM}, Multiple Imputation \cite{MI}, and Data Augmentation \cite{DA}.


As anticipated, a remarkable feature of this method relates to the possibility of estimating the effect of fluctuations 
on the shapes of PDFs that correspond to particles originating from a signal hard parton scattering as opposed to low-energy QCD background. The shapes of the particle-level signal and background PDFs in a given data set can in fact differ notably from those of the corresponding templates obtained from high-statistics control samples, where the effect of fluctuations on the PDF shapes is normally averaged out. As expected, the deviation of the actual PDF shapes from the shapes of the corresponding control sample templates in general becomes more and more notable as the number of particles in the input data set is reduced.


The algorithm estimates the shapes of the PDFs corresponding to particles associated with a signal hard parton scattering as opposed to soft QCD background. This is done by iteratively mapping particles to signal or background using the data to refine initial conditions obtained from the control samples. 
The effect of fluctuations on the PDF shapes is encoded in the stationary distribution of the Markov Chain, the existence and uniqueness of which is discussed in \cite{gibbshep2, gibbshep}. 
%



\begin{figure*}
\begin{minipage}{20pc}
  \centering
  \subfloat[]{
  \includegraphics[scale=0.27]{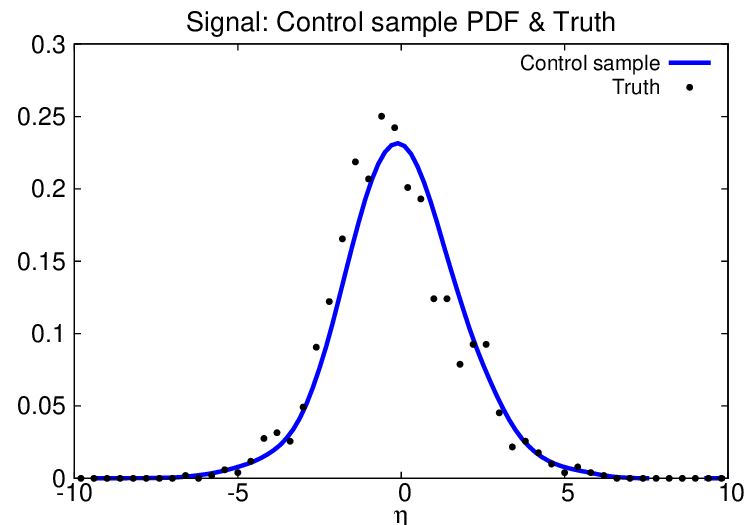} 
  }
  \subfloat[]{
  \includegraphics[scale=0.27]{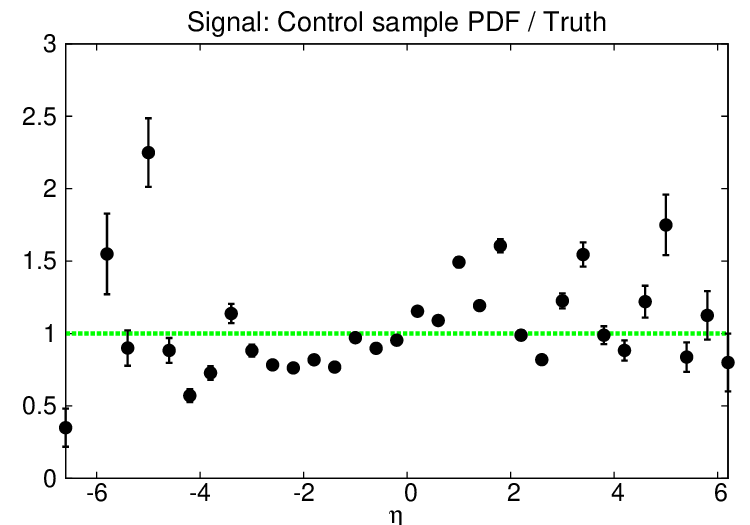} 
  }
  \\
  \subfloat[]{
  \includegraphics[scale=0.27]{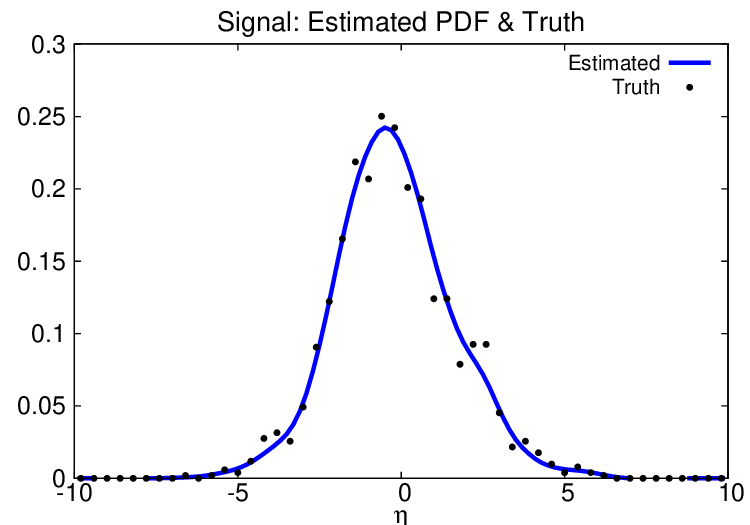} 
  }
  \subfloat[]{
  \includegraphics[scale=0.27]{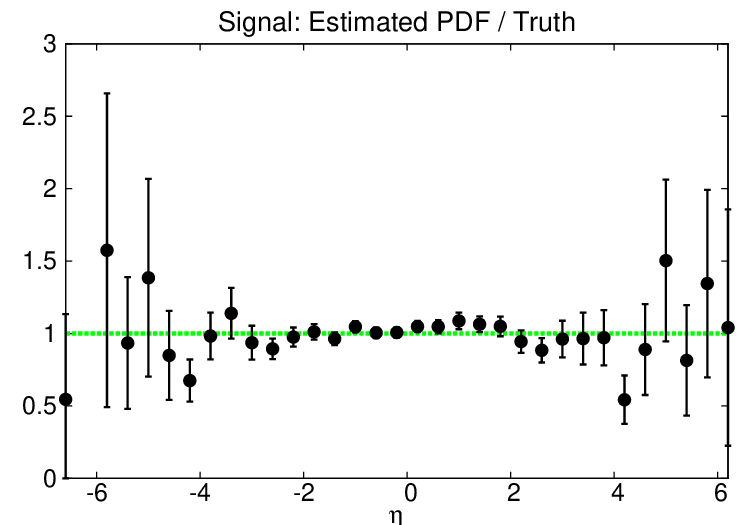} 
  }
\end{minipage}
\begin{minipage}{18pc}
  \centering
  \subfloat[]{
  \includegraphics[scale=0.25]{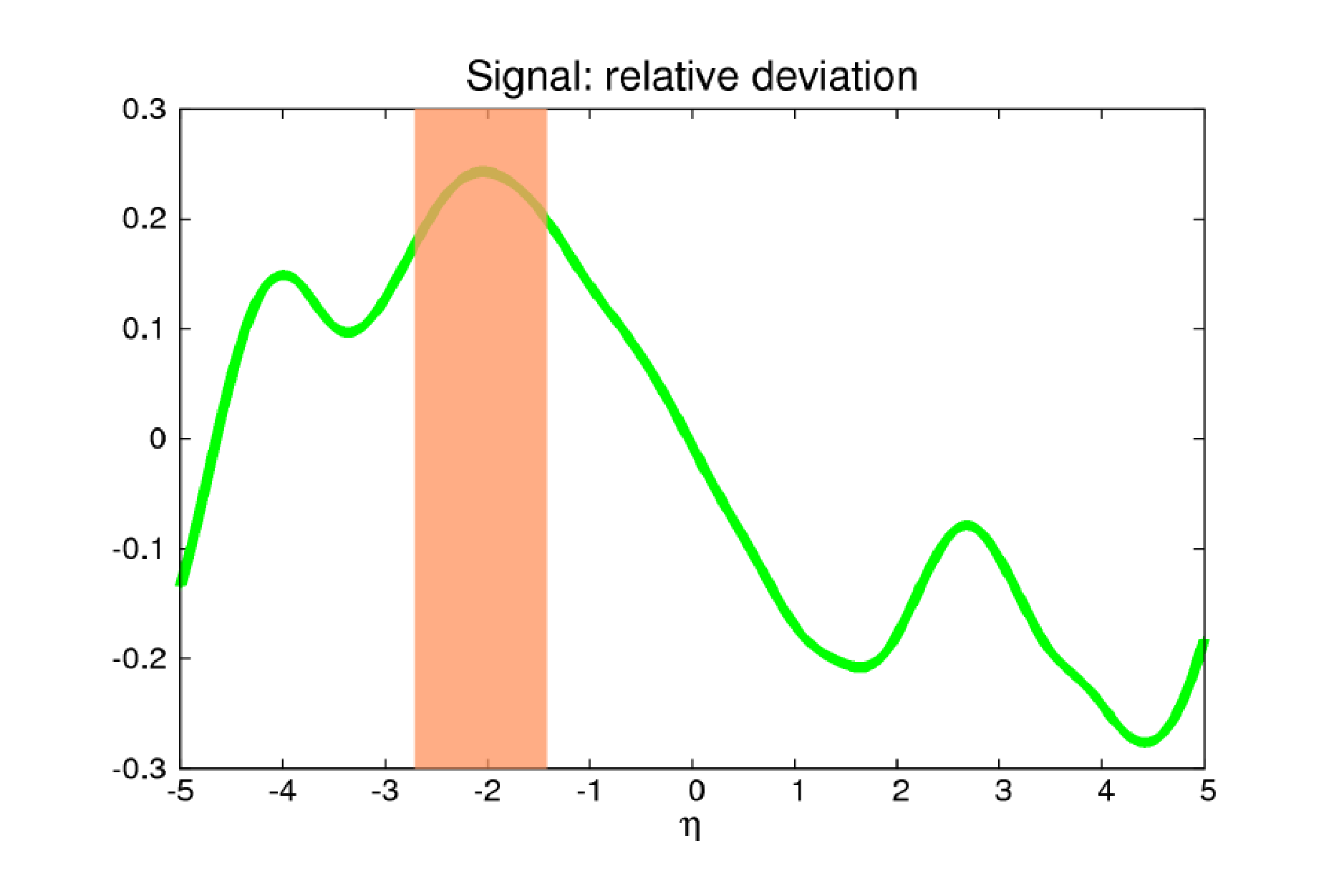} 
  }
\end{minipage}
\begin{minipage}{38pc}
  \caption{\label{fig:compar}
(a) Monte Carlo true signal particle $\eta$ PDF (points), superimposed with the corresponding control sample template (curve) \cite{gibbshep2}. The plot highlights the effect of fluctuations on the PDF shape, $\chi^2/ndof = 38.9$. (b) Ratio between control sample and true PDF corresponding to (a). (c) The same true distribution (points) superimposed with the PDF estimated by the algorithm (curve), $\chi^2/ndof = 0.98$. (d) Ratio between estimated and true PDF corresponding to (c). (e) Relative difference between estimated and control sample PDF. For the sake of illustration, the vertical band represents a hypothetical jet around $\eta=-2$.}
\end{minipage}
\end{figure*}

Figure \ref{fig:compar} (a) displays the true 
$\eta$ distribution of particles from Monte Carlo $gg\rightarrow t\bar{t}$ normalised to unit area (points), superimposed with the PDF template obtained from a high-statistics control sample (curve) \cite{gibbshep2}. The figure shows how the true distribution deviates from the control sample template due to the presence of fluctuations in the data. Figure \ref{fig:compar} (c) shows the same true distribution (points) superimposed with the PDF estimated using the algorithm (curve). The agreement with the true distribution is remarkably improved, corresponding to $\chi^2/ndof=0.98$ as opposed to $\chi^2/ndof=38.9$ from figure \ref{fig:compar} (a). 
The corresponding ratios are given in figures \ref{fig:compar} (b) and (d). 


Since the data set used in \cite{gibbshep2} comprises a number of particles that is in line with typical LHC 
event 
multiplicities, it makes sense to use those results to illustrate the anticipated performance of the algorithm on individual LHC events. 
Figure \ref{fig:compar} (e) shows the difference between the signal $\eta$ PDF estimated by the algorithm and the control sample template, normalised to the latter. The vertical band corresponds to a hypothetical jet around $\eta=-2$. Given that the relative difference between the actual PDF and the control sample template 
can be as high as 20\% in that interval of $\eta$, if one were to map individual particles inside such a hypothetical jet to signal or background using the control sample PDF, i.e. neglecting the effect of fluctuations, the number of signal particles inside the jet would be underestimated by as much as 20\%. 
%
For this reason, this technique can also 
be used to obtain precise estimates of the fraction of soft QCD particles 
inside individual jets, thereby taking into account the effect of fluctuations at the particle level.

Finally, with regard to execution time, the algorithm processed the Monte Carlo data sets used in \cite{gibbshep2} in $\sim20~\mbox{s}$ on a 2~GHz Intel Processor with 1~GB RAM without any optimisation. We consider such performance reasonable for offline use.

\section{Outlook}
\label{outlook}

Our hope is that this new approach will complement existing techniques for subtraction of low-energy QCD background at hadron colliders. In fact, since it is based on a different principle and it works in a different way as compared to state-of-the-art techniques, we expect it to further improve physics analysis in high-luminosity hadron collider environments when used in combination with existing methods.
%
%
We anticipate that particle-level event filtering will provide a more significant contribution as pileup rates and average event multiplicities increase, e.g. to improve jet mass and $\slashed{p}_T$ resolution, depending on the analysis. 
%
We also expect the ability to reject soft QCD contamination particle by particle thereby taking into account the effect of fluctuations on the 
PDF shapes inside individual events to further improve background subtraction inside fat jets 
from decays of possible new heavy particles. 




\section{Conclusions}
\label{concl}

We have 
discussed 
the potential of our algorithm \cite{gibbshep2} to implement a novel particle-by-particle filtering procedure for individual events at hadron collider experiments, which we envisage as a possible new data processing stage upstream of the existing jet reconstruction pipelines. One central aspect is the possibility to map individual particles to a hard scattering as opposed to low-energy QCD background, thereby taking into account the effect of particle-level fluctuations on the shapes of signal and background PDFs inside individual events.

We have shown that, if one is to map individual particles inside events to a hard parton scattering as opposed to soft QCD interactions, using PDFs obtained from independent high-statistics control samples does not take into account the effect of fluctuations, and can lead to a shift in the estimated number of signal particles as high as 20\%. On the other hand, the particle-level PDFs estimated using our algorithm were found to be in remarkable agreement with the true distributions on the Monte Carlo data sets analysed in \cite{gibbshep2}. This method can therefore also produce precise estimates of the fraction of soft QCD particles inside individual jets.
%



Our hope is that this approach will 
improve the resolution of jet observables in high-luminosity environments when used in combination with state-of-the-art techniques such as jet grooming algorithms, 
e.g. with reference to the mass of fat jets from boosted decays of possible new heavy particles. 
More generally, it is our opinion that particle-by-particle filtering of individual events based on high-precision particle-level PDFs has the potential to become a useful ingredient of physics analysis at future high-luminosity hadron collider experiments.







\section{Acknowledgments}
The author wishes to thank the High Energy Physics Group at Brunel University for a stimulating environment, and particularly Prof. Akram Khan and Prof. Peter Hobson. Particular gratitude also goes to the High Energy Physics Group at University College London, especially to Prof. Jonathan Butterworth for his valuable comments. The author also wishes to thank Prof. Trevor Sweeting at the UCL Department of Statistical Science, as well as Dr. Alexandros Beskos at the same department for fruitful discussions. Finally, particular gratitude goes to Prof. Carsten Peterson and to Prof. Leif Lönnblad at the Department of Theoretical Physics, Lund University.


\end{multicols}
\end{document}